\documentclass{aa}  

\usepackage{caption}
\usepackage{graphicx}
\usepackage{ulem}
\usepackage{stfloats} 
\usepackage{txfonts}
\usepackage{color}
\definecolor{linkcolor}{rgb}{0,0,.7}
\definecolor{oldText}{rgb}{0.5,0,0}
\definecolor{newText}{rgb}{0,.5,0}

\usepackage[colorlinks=true]{hyperref}
\hypersetup{
  colorlinks   = true,
  urlcolor     = linkcolor,
  linkcolor    = linkcolor, 
  citecolor   = linkcolor 
}

\renewcommand{\H}{\text{H}}
\renewcommand{\O}{\text{O}}
\renewcommand{\doi}[1]{\textcolor{linkcolor}{\href{http://doi.org/#1}{\textbf{doi}}}}
\newcommand{\ads}[1]{\textcolor{linkcolor}{\href{#1}{\textbf{ads}}}}
\renewcommand{\arxiv}[1]{\textcolor{linkcolor}{\href{https://arxiv.org/abs/#1}{\textbf{arXiv}}}}

\newcommand{\Fe}{\text{Fe}}
\newcommand{\Si}{\text{Si}}
\newcommand{\Mg}{\text{Mg}}

\renewcommand{\O}{\text{O}}

\renewcommand{\H}{\text{H}}

\renewcommand{\S}{\text{S}}

\newcommand{\g}{\text{(g)}}
\newcommand{\s}{\text{(s)}}

\newcommand{\e}[1]{\cdot 10^{#1}}

\begin{document} 

   \title{The effects of transport processes on the bulk composition of the first generation of planetesimals interior to the water iceline}
 \titlerunning{The effects of transport processes on planetesimal composition}
 
   \author{M. Oosterloo
          \inst{1,2}
          \and
          I. Kamp
          \inst{1}
          \and
          W. van Westrenen
          \inst{2}
          }

   \institute{Kapteyn Astronomical Institute, University of Groningen, Landleven 12, 9747 AD Groningen, The Netherlands
                  \and
               Department of Earth Sciences, Vrije Universiteit Amsterdam, De Boelelaan 1085,
1081 HV Amsterdam, The Netherlands
               }   

   \date{Received: 8.10.2024; accepted: 11.4.2025}

  \abstract
   {The CHNOS elemental budgets of rocky planets are crucial for their structure, evolution and potential chemical habitability. It is unclear how the nonlocal disk processes affecting dust in planet-forming disks affect the CHNOS elemental budgets of nascent planets both inside and outside the Solar System.}
   {We aim to quantify the coupled effect of dynamical and collisional processes on the initial refractory CHNOS budgets of planetesimals, forming interior to the water ice line for a Solar and non-Solar composition consistent with the star HIP 43393.}
   {We utilize the SHAMPOO code to track the effects of dynamical and collisional processes on 16000 individual dust monomers. Each monomer is here assigned a refractory chemical composition and mineralogy informed by the equilibrium condensation code GGCHEM given the P-T conditions at the initial position of the monomer. Monomers travel embedded in aggregates through a young class I disk, whose structure is calculated with the ProDiMo code. Furthermore, monomers are allowed to undergo dehydration and desulfurization.}
   {We find that solid material becomes well-mixed both radially and vertically. For both the Solar and HIP 43393 compositions, the solid phase in the disk midplane regions interior to $r\eqsim 0.7$ AU can become enriched in hydrogen and sulfur by up to 10 at$\%$ relative to predictions from purely local calculations. This originates from the inward radial transport of hydrated and sulfur-bearing minerals such as lizardite and iron sulfide.}
   {Nonlocal disk processing in a young turbulent, massive disk can lead to significant compositional homogenization of the midplane dust and by extension of the initial composition of planetesimals. Planetesimals forming at $r<0.7$ AU may become enriched in hydrated minerals and sulfur, which could result in more widespread aqueous alteration interior to the water iceline compared to planetesimals that emerge from more locally processed dust.}

    \keywords{}

    \maketitle

    \section{Introduction}
    \label{sec:3.1}

The past years have seen significant progress in the observational and numerical characterization of the diversity in rocky exoplanets in terms of their size, mass, and surface and interior composition. Of particular interest are the planetary elemental budgets of carbon, hydrogen, nitrogen, oxygen and sulfur (CHNOS) due to their importance for the chemical habitability of rocky exoplanets. 
This importance stems from their prevalence in bio-essential molecules and their effects on the structure and mineralogy of planetary interiors \citep{hakim_mineralogy_2019, hoehler_lifes_2020, krijt_chemical_2022, jorge_forming_2022}. 

The amounts of CHNOS in rocky planets are closely connected to the elemental abundances of the host star since they form from the same molecular cloud. Rocky planets form from the dust grains present in the inner regions of the protoplanetary disk surrounding the young host star \citep[e.g.][]{armitage_astrophysics_2010,raymond_planet_2022, drazkowska_planet_2023}. Evidence from white dwarfs that are polluted from accreting exoplanetary material suggests that exoplanetary compositions are similar to stellar abundances \citep{bonsor_host-star_2021}. At least some of these rocky exoplanets may have a bulk composition and mineralogy similar to Earth \citep{doyle_oxygen_2019}. 

Spectroscopic surveys of stellar photospheric elemental abundances have demonstrated considerable variations in the elemental abundances in stars in the solar neighbourhood \citep{hinkel_stellar_2014, buder_galah_2018}. In particular, peculiar mineralogies could emerge in planets forming around stars with a super-solar C/O or S/Fe ratio \citep{hakim_mineralogy_2019, jorge_forming_2022, oosterloo_interiors_2024}. A commonly used method to explore the implications of a given stellar elemental abundance for planetary mineralogies are equilibrium condensation models such as CONDOR, \citep{lodders_lanthanide_1993}, GGCHEM \citep{woitke_equilibrium_2018} and FASTCHEM \citep{stock_fastchem_2018, kitzmann_fastchem_2024}. These models have provided valuable insights in the range of bulk rocky exoplanet compositions in the Solar neighbourhood \citep{jorge_forming_2022, spaargaren_plausible_2023}. Also, bulk Earth elemental abundances of Mg, Si, Fe, O and S \citep{mcdonough_terrestrial_2021} have been reproduced with equilibrium condensation models surprisingly well \citep{jorge_forming_2022}. 

Despite these results, translating a stellar elemental abundance into planetary elemental abundances is a highly nontrivial endeavour. Observational evidence suggests that planets readily start to form in class I disks \citep[e.g.][]{sheehan_multiple_2018, segura-cox_four_2020}, which are dynamically active environments, where stellar outbursts due to accretion may result in significant re-condensation of dust in the inner disk regions \citep[e.g.][]{fischer_accretion_2022} and ongoing infall can contribute fresh material to disks in class 0/I stages \citep[e.g.][]{hueso_evolution_2005, gieser2024}. In this environment, the first planetesimals are thought to form over periods of $10^5$ yr from the dust grains undergoing collisional and dynamical processing \citep[e.g.][]{kleine_hf-w_2009, drazkowska_planet_2023}. This growth is by no means smooth, with dust aggregate fragmentation, bouncing and radial drift providing various barriers against formation of planetesimals through collisional growth \citep{weidenschilling_aerodynamics_1977, nakagawa_settling_1986, blum_experimental_1993, zsom_outcome_2010, birnstiel_simple_2012, drazkowska_planet_2023}. Moreover, transport processes such as drift, settling and turbulent diffusion can result in significant displacement and redistribution of planetesimal-forming dust throughout the protoplanetary disk in the $10^5$ yr time frame \citep[e.g.][]{armitage_astrophysics_2010, ciesla_residence_2010, ciesla_residence_2011, piso_co_2015, oosterloo_shampoo_2023, oosterloo_effect_2024}. 

An important limitation of equilibrium condensation models arises in this context since condensation sequences are generated for a specific P-T profile. In reality, the dust populations in protoplanetary disks that ultimately give rise to planetesimals and planets possess a significant nonlocal component, in particular in the inner disk regions \citep[$r<10$ AU,][]{raymond_planet_2022, timmermann_revisiting_2023, oosterloo_effect_2024}. This component originates from diffusion and drift processes, that may transport individual dust grains that have condensed under certain P-T conditions over vast distances in the disk \citep{ciesla_residence_2010, ciesla_residence_2011, raymond_planet_2022}. Which of these transport processes dominates the dynamical behaviour of individual dust grains depends on the grain size, which is a product of the collisional history of the dust grain. Previously, the coupling between dynamical, collisional and ice processing for the volatile CHNOS budgets in the outer disk has sparked the development of the SHAMPOO code \citep{oosterloo_shampoo_2023, oosterloo_effect_2024}. However, significant amounts of the total CHNOS elemental budgets are incorporated in a more refractory reservoir \citep[e.g.][]{kama_abundant_2019, oberg_astrochemistry_2021, jorge_forming_2022}. 

In this work, we aim to focus on the evolutionary phase just predating the formation of planetesimals, so a young massive disk. For simplicity, 
we will neglect the potential infall of fresh material. The goal of our work is to quantify the coupled effects of dynamical processes (turbulence-driven diffusion, radial drift and vertical settling) and collisional processes (coagulation, fragmentation and erosion) on the refractory CHNOS elemental budgets of refractories interior to the water ice line. 
To cover a range of potential mineralogies identified in earlier works (see above), we explore solar and super-solar sulfur abundance. We use 
the SHAMPOO\footnote{\url{https://github.com/moosterloo96/shampoo}} code \citep{oosterloo_shampoo_2023}, a stochastic model that tracks the effects of dynamical, collisional and ice processing on individual dust monomers travelling throughout a static disk environment. This disk environment is here calculated using the thermochemical disk code ProDiMo \citep{woitke_radiation_2009, kamp_radiation_2010, thi_radiation_2011, thi_radiation_2013}, while dust is assigned refractory compositions assuming equilibrium condensation conditions at the local temperature and pressure with the GGCHEM code \citep{woitke_equilibrium_2018}. 

In Sect. \ref{sec:3.2} we explain our methodology and assumptions regarding the protoplanetary disk environment, and equilibrium condensation and briefly review the ProDiMo, GGCHEM and SHAMPOO codes. In Sect. \ref{sec:3.3} we explore the nonlocal behaviour of dust in the inner disk and its effects on the refractory CHNOS budget of local dust populations, and we discuss their implications for planet composition in Sect. \ref{sec:3.4}.

    \section{Methods}
    \label{sec:3.2}

\subsection{Disk structure}
\label{sec:4.2.1}

To model the behaviour of dust monomers in SHAMPOO, we require the gas and dust mass densities $\rho_\text{g}$ and $\rho_\text{d}$ and temperatures $T_\text{g}$ and $T_\text{d}$ as a function of radial and vertical position $r,z$ throughout the protoplanetary disk. These disk density and temperature structures are calculated on a pre-specified grid of radial and vertical positions $r_i,z_j$ using the thermo-chemical disk code ProDiMo \citep[][]{woitke_radiation_2009, kamp_radiation_2010, thi_radiation_2011, thi_radiation_2013}. In order to obtain these quantities at any position $r,z$, we utilize linear interpolation.

This work considers the \texttt{vFrag1} background disk model from \cite{oosterloo_shampoo_2023}, which comprises a smooth, axisymmetric 0.1$M_\odot$ disk around a 0.7$M_\odot$ T Tauri star, with disk parameters consistent with a young class I disk. The disk dust-to-gas mass ratio is 0.01 and the inner and outer disk radii are 0.07 AU and 600 AU, respectively. The maximum grain size $a_\text{max}$ is assumed to be limited by fragmentation \citep[][]{birnstiel_simple_2012}, with $a_\text{max}(r)$ being consistent with a fragmentation velocity of $v_\text{frag}=1$ m\,s$^{-1}$. A full table of model parameters and visual overviews of $\rho_\text{g}$, $\rho_\text{d}$, $T_\text{g}$ and $T_\text{d}$ are provided in \cite{oosterloo_shampoo_2023}. In addition to these quantities, the background disk model also provides the size distribution for dust grains of size $a$, $f(r,z,a)\propto a^{-a_\text{pow}}$, with $a_\text{pow}=3.5$. The background dust size distribution is vertically distributed according to settling effects using the methodology outlined in \cite{woitke_consistent_2016, oosterloo_shampoo_2023}. For the radial disk density structure, we use a power law profile for the disk column density. The power law index is here set equal to unity, such that the column density as a function of radial distance $r$ decreases as $\Sigma(r) \propto r^{-1}$. An exponential tapering applies for radii larger than 100 AU \citep[see also][]{woitke_consistent_2016, oosterloo_shampoo_2023}. The vertical gas density structure is described by a Gaussian. The dust scale height for dust grains, which can differ for larger dust grains that are dynamically decoupled from the gas is calculated according to \cite{dubrulle_dust_1995}. Turbulence is described with a single value $\alpha=10^{-3}$, parametrizing the turbulence strength \citep[][]{shakura_black_1973}. It was found in \cite{oosterloo_effect_2024} that for the specific background model considered in this work, the Stokes number for the aggregates that contain most dust mass never exceeds St = $10^{-3}$ interior to $r=10$ AU. Therefore, diffusion driven by turbulence is the dominant transport process for $r<1.5$ AU. Our disk setup hence corresponds more closely to an evolutionary stage just predating the efficient formation of planetesimals. The temperature structures for the gas and dust, $T_\text{g}$ and $T_\text{d}$ are obtained from the local radiation field $J_\nu$, which results from solving the local continuum radiative transfer. More elaborate discussions on specific heating and cooling processes considered in ProDiMo are presented in \cite{woitke_radiation_2009, woitke_consistent_2016, thi_radiation_2011, aresu_x-ray_2011, oberg_circumplanetary_2022}. In this particular disk model, the ice line of water is located at $r=2$ AU in the disk midplane, with the mass of water adsorbed on dust becoming negligible compared to the total dust mass between $r=1.5$ AU and $r=2$ AU \citep{oosterloo_effect_2024}. We here define the ice line of water as the location where more than 50$\%$ of the total amount of water becomes incorporated in dust grains as ice.

\subsection{Dust dynamics}
\label{sec:4.2.2}

SHAMPOO is a stochastic model that tracks the effects of dynamical, collisional and ice processing on individual dust monomers which represent units of dust mass \citep{oosterloo_shampoo_2023}. For this study, we focus on the dynamical and collisional processing of dust. Individual dust monomers have a fixed radius $s_\text{m}=5\e{-8}$ m, while collisional growth usually results in dust monomers being incorporated in larger aggregates. In this study, we focus on the disk region inside $r=1.5$ AU, such that the contribution of volatile CHNOS-bearing ices to the total solid-phase CHNOS elemental budgets is small compared to refractories \citep{oosterloo_effect_2024}, and thus ice processing and the depth $z_\text{m}$ at which dust monomers are incorporated inside an aggregate can be ignored \citep{oosterloo_shampoo_2023}. For dynamical transport, SHAMPOO calculates the vertical settling and radial drift for the monomer inside its aggregate of radius $s_\text{a}$, while turbulent diffusion is calculated using a random-walk approach akin to \cite{ciesla_residence_2010, ciesla_residence_2011}. Collisions are treated in a stochastic, collision-by-collision fashion, following the approach of \cite{krijt_dust_2016}, with collisional outcomes comprising coagulation, fragmentation and erosion. These result in temporal variations in the aggregate size $s_\text{a}$ and thus the dynamical behaviour of the dust monomer. Furthermore, dynamical transport can affect collisional processing due to its effects on the collision rates via the position-dependence of the local dust size distribution $f(r,z,a)$. An example of a resulting evolutionary monomer trajectory is shown in Appendix \ref{sec:4.AA}, while more comprehensive interpretations of monomer trajectories that include the evolution of volatiles are provided in \cite{oosterloo_shampoo_2023}.

We perform a similar analysis as outlined in \cite{oosterloo_effect_2024}. We consider the evolutionary trajectories of 16 000 dust monomers over 100 kyr with the parameter set presented in \cite{oosterloo_shampoo_2023}. Trajectories of individual monomers are treated separately to allow for both parallel and sequential computational evaluation. The initial positions of the monomers are chosen randomly from a log uniform and uniform distribution in $r\in [0.08, 5]$ AU and $z/r\in [-0.1, 0.1]$, respectively. The initial aggregate sizes $s_\text{a}$ are determined from the mass-weighted local dust size distribution. The trajectories of individual monomers comprise many timesteps $r_n$, $z_n$, with $n\in[0,N_t]$ and $N_t$ being the total number of timesteps in the evolutionary trajectory of a monomer. In order to derive the properties of local dust populations at fixed positions in the protoplanetary disk, we utilize the discrete coordinate grid in the background model $r_i$, $z_j$, and assign different monomer timesteps to their nearest background model coordinate. Via this method, properties of local dust can be derived at each background model grid point $r_i$, $z_j$ via the average properties of many (typically more than $10^5$) monomer timesteps. We also weigh each monomer timestep utilizing the scheme outlined in \cite{oosterloo_effect_2024}. Each monomer is labelled with a static refractory composition representative for the temperature and pressure conditions under which it formed.

\subsection{Monomer composition}
\label{sec:4.2.3}

\begin{table}[]
\caption{Elemental atomic abundances used in GGCHEM for Solar composition and the sulfur-rich composition associated with HIP 43393. Abundances are given in log units and are normalized to silicon ($\log n_\Si=9$). Solar abundances are taken from \cite{asplund_chemical_2009}, while the abundances of HIP 43393 are from the Hypatia catalogue \citep{hinkel_stellar_2014}.}
\begin{tabular}{r|l|l|r|l|l}
\hline
{Elem.} & {Solar} & {HIP 43393} & {Elem.} & {Solar} & {HIP 43393} \\\hline\hline
H & 13.49 & 13.82 & He & 12.42 & 12.75\\
Fe & 8.99 & 8.75 & C & 9.92 & 9.94\\
O & 10.18 & 10.54 & Mg & 9.09 & 9.13\\
\textbf{Si} & \textbf{9.00} & \textbf{9.00} & Ca & 7.83 & 7.72\\
Ti & 6.44 & 6.56 & Li & 2.54 & 2.56\\
N & 9.32 & 9.34 & F & 6.05 & 6.07\\
Na & 7.73 & 7.60 & Al & 7.94 & 8.01\\
P & 6.90 & 6.92 & Cl & 6.99 & 7.01\\
K & 6.52 & 6.54 & V & 5.42 & 5.48\\
Cr & 7.13 & 6.97 & Mn & 6.92 & 6.52\\
Ni & 7.71 & 7.48 & Zr & 4.07 & 4.20\\
S & 8.61 & 9.04 & W & 2.34 & 2.36\\\hline
\end{tabular}
\label{tab:abundances} 
\end{table}

Class I disks are dynamic environments, with the central star undergoing regular accretion outbursts \citep[e.g.][]{fischer_accretion_2022}. These outbursts give rise to evaporation and re-condensation of refractory material under local disk equilibrium conditions and could help to explain differences between the elemental compositions of interstellar grains and rocky inner Solar System bodies \citep[e.g.][]{bergin_tracing_2015, anderson_destruction_2017, jorge_forming_2022}{}. From these considerations, we assume that dust monomers have formed under local equilibrium conditions and calculate monomer compositions using the GGCHEM code.

The GGCHEM code was originally developed by \cite{gail_primary_1986} and completely rewritten by \cite{woitke_equilibrium_2018}. It determines the chemical composition of gases in thermo-chemical equilibrium along with the formation of equilibrium condensates down to a temperature of 100 K via Gibbs free energy minimalization \citep[e.g.][]{white_chemical_1958, eriksson_thermodynamic_1971}{}{}. Thermodynamical data is taken from the NIST-JANAF \citep[][]{chase_janaf_1982, chase_janaf_1986}{} and the SUPRCRTBL databases \citep[]{johnson_supcrt92_1992, zimmer_supcrtbl_2016}{}{}. For a full summary of the numerical methodology and the calculation of condensate phases in the GGCHEM code, we refer the interested reader to \cite{woitke_equilibrium_2018}. In GGCHEM, 24 elements can be included in the calculation of the condensation sequence (see Table \ref{tab:abundances}). These elements can form 552 different chemical species, including many charged species. Furthermore, 245 different condensates can form in GGCHEM. For this study, the condensate set in GGCHEM was expanded to include several sulfate molecules (CoSO$_4$, CuSO$_4$, Fe$_2$(SO$_4$)$_3$, FeSO$_4$ and MgSO$_4$). Thermodynamic data for these species was taken from the NIST-JANAF database, where we utilize the fitting coefficients reported by \cite{kitzmann_fastchem_2024}. 

In this work, we use GGCHEM to determine the composition of monomers based on the position $r, z$ where the monomers are initialized in SHAMPOO. Each monomer is labelled with a unique composition obtained from a condensation sequence calculated in GGCHEM, down to the ambient gas temperature and pressure $T_\text{g}, P$ at $r, z$. During the calculation of each condensation sequence, the temperature is gradually decreased from 2500 K to the local gas temperature $T_\text{g}$, while the pressure is kept fixed at the local gas pressure. This leads to an inconsistency between the dust composition used in the opacities of the background disk model and the one from GGCHEM \citep[see][for a more consistent treatment in the context of CAI formation]{woitke_cai_2024}, but it enables a clean assessment of the importance of the dynamical and collisional processing of dust aggregates.

We consider condensation sequences for Solar composition and for a sulfur-rich stellar composition. The latter was found to have implications for the mineralogy of condensates and potentially planet structure \citep[][]{jorge_forming_2022, oosterloo_interiors_2024}{}. The photospheric composition of HIP 43393 was chosen based on the mineralogy of condensates previously derived for protoplanetary disk conditions around this star by \cite{jorge_forming_2022} and the previous exploration of this composition in laboratory experiments under P-T conditions typical for planetary interiors \citep[][]{oosterloo_interiors_2024}{}. The elemental abundances for GGCHEM are shown for both the Sun and HIP 43393 in Table \ref{tab:abundances}.

    \section{Results}
    \label{sec:3.3}

\subsection{Nonlocal dust in the inner disk}
\label{sec:4.3.1}

\begin{figure}
    \centering
    \includegraphics[width=.48\textwidth]{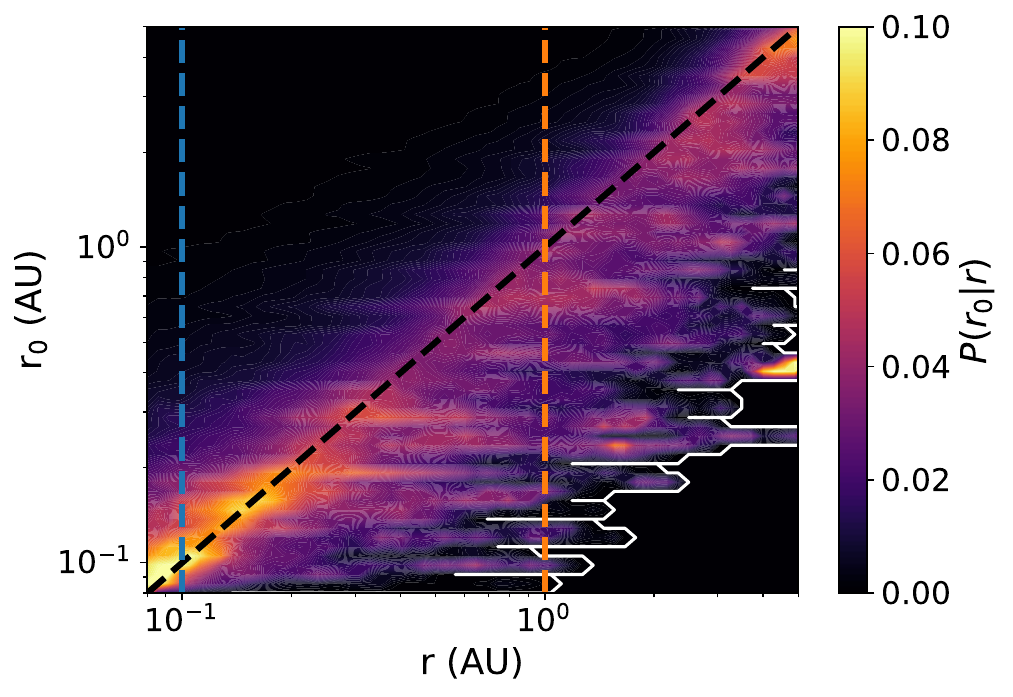}
    \caption{Overview of the distribution $P(r_0|r)$ of the monomer positions of origin $r_0$ as a function of radial position $r$. The 1:1 axis is shown in black dashes, while the coloured dashed lines indicate the radial locations $r=0.1, 1$ AU of the slices shown in Fig. \ref{fig:31NonLocalBiqSolarOriginDiagramSnapshots}.}
    \label{fig:31NonLocalBiqSolarOriginDiagram}
\end{figure}
\begin{figure}
    \centering
    \includegraphics[width=.48\textwidth]{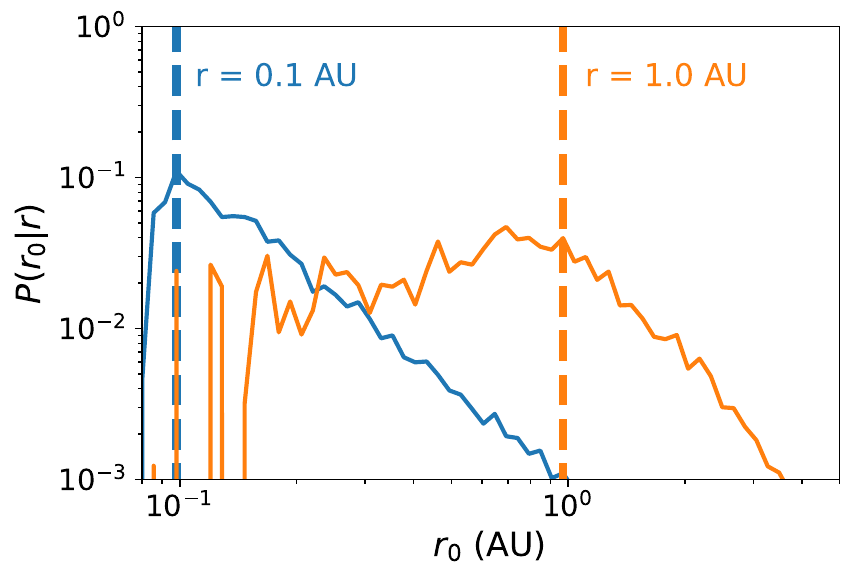}
    \caption{Slices of $P(r_0|r)$ from Fig. \ref{fig:31NonLocalBiqSolarOriginDiagram} as a function of monomer position of origin $r_0$ at $r=0.1$ and $r=1$ AU.}
    \label{fig:31NonLocalBiqSolarOriginDiagramSnapshots}
\end{figure}
\begin{figure*}[ht!]
    \centering
    \includegraphics[width=\textwidth]{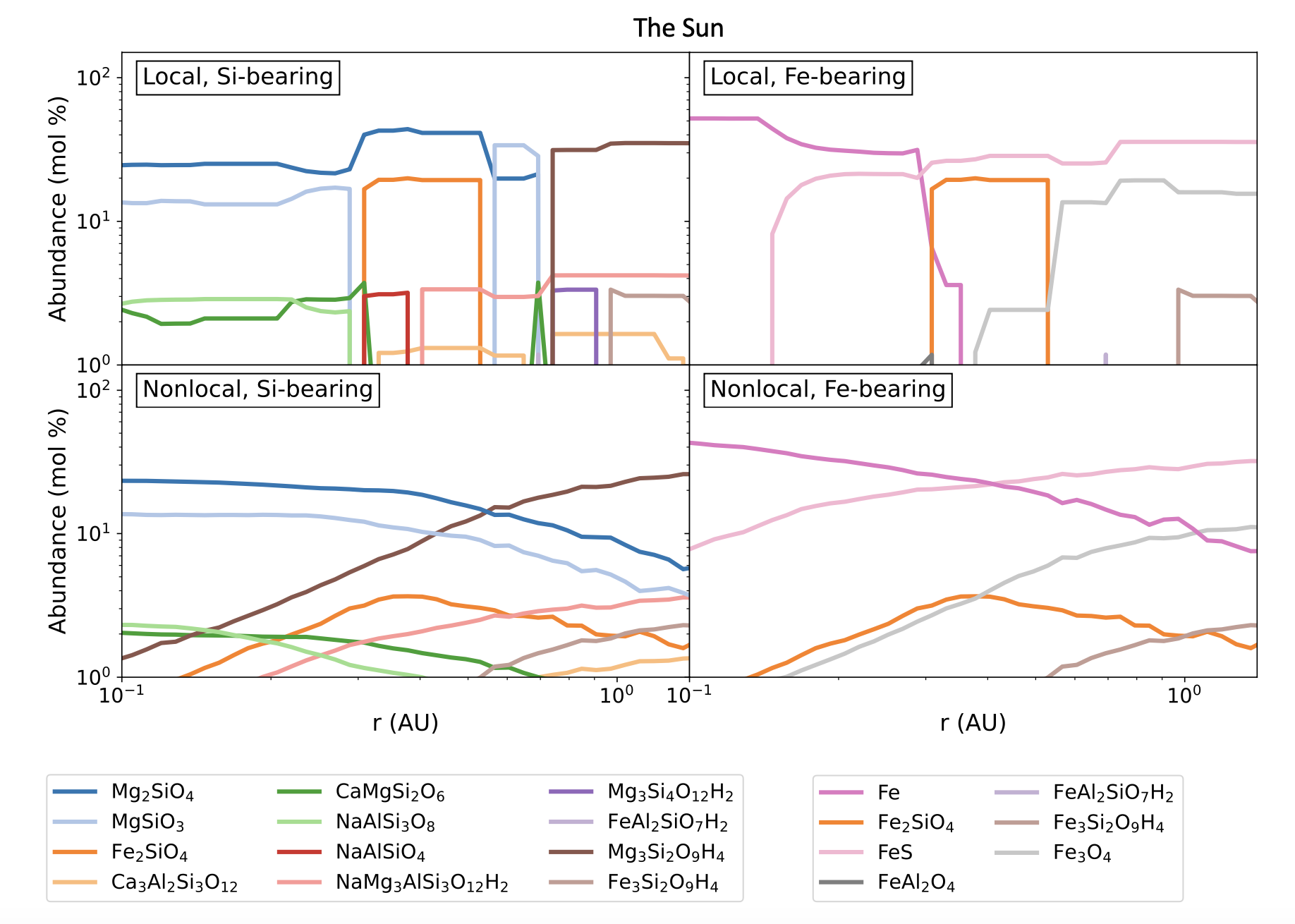}
    \caption{Condensate composition as a function of the radial position of silicate (left column) and metallic (right column) minerals for a Solar nebular composition. The top row depicts the mineralogy expected from equilibrium condensation under the local temperature and pressure conditions (initial conditions). The bottom row is the nonlocal composition obtained from monomers undergoing radial and vertical transport and collisional processing after 100 kyr.}
    \label{fig:32NonLocalBiqSolarSolarMineralogy}
\end{figure*}
\begin{figure*}[ht!]
    \centering
    \includegraphics[width=\textwidth]{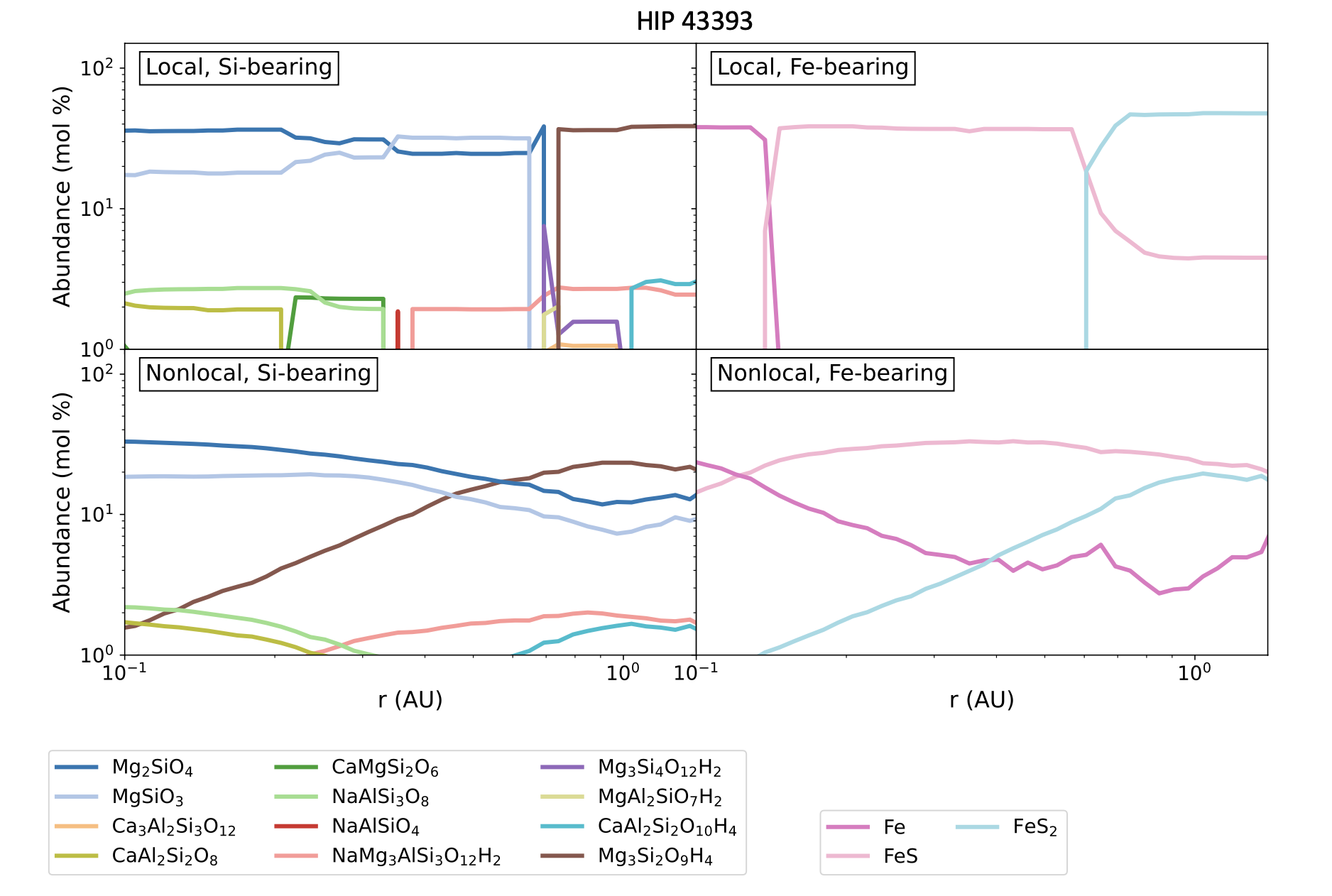}
    \caption{Condensate composition as a function of the radial position of silicon-bearing (left column) and iron-bearing (right column) minerals for a nebular composition consistent with the photosphere of HIP 43393. As in Fig. \ref{fig:32NonLocalBiqSolarSolarMineralogy}, the top and bottom row denote the local initial conditions and nonlocal compositions after 100 kyr, respectively.}
    \label{fig:32NonLocalBiqSolarStar4Mineralogy}
\end{figure*}

As already found in \cite{oosterloo_shampoo_2023}, diffusion driven by turbulence is the dominant transport process in young, massive disks. This is even more the case for the inner regions of these disks, $r\lesssim 1.5$ AU, where gas densities are high. In Fig. \ref{fig:31NonLocalBiqSolarOriginDiagram} we explore the distributions $P(r_0|r)$ of radial origin $r_0$ of each monomer that has visited the radial position $r$. $P(r_0|r)$ is here the probability of encountering a unit of mass originating from $r_0$ in the dust at position $r$. Note that this means that individual monomers contribute to the distributions at different radial positions at different times in their trajectories. It is clear from Fig. \ref{fig:31NonLocalBiqSolarOriginDiagram} that dust in the inner disk is even better mixed than at the larger radial positions explored in \cite{oosterloo_effect_2024}. At $r\,\lesssim\,$2 AU, the distribution of $r_0$ becomes increasingly more skewed towards positions of origin farther out than $r$ as the inner boundary of the monomer sampling interval at 0.08~AU is approached. At larger radial positions $r$, the distribution in $r_0$ becomes more evenly spread out towards locations both further in and out. Fig. \ref{fig:31NonLocalBiqSolarOriginDiagramSnapshots} shows that over the period of $10^5$ yr, material from 1 AU can diffuse inwards towards 0.1 AU, close to the inner disk edge, while the reverse is also true. Altogether dust in the inner disk can become significantly mixed via turbulent diffusion on timescales comparable to or shorter than the typically derived planetesimal formation timescale, $\sim\,10^5$ yr \citep[e.g.][]{kleine_hf-w_2009, drazkowska_planet_2023}.

\subsection{Effects on mineralogy}
\label{sec:4.3.2}

As a next step, we compare the solid phase mineralogy in the disk midplane under the influence of dynamical and collisional processes (after 100 kyr) to fully local equilibrium condensation (initial conditions). Fig. \ref{fig:32NonLocalBiqSolarSolarMineralogy} compares the results for a Solar composition of a fully local GGCHEM condensation sequence under the local P-T conditions in the disk midplane as a function of radial position $r$ to the average composition derived from all monomer timesteps at the same radial position. We here focus on the region between 0.1 and 1.5 AU, where no significant amounts of water ice form in either the background model or GGCHEM.

From Fig. \ref{fig:32NonLocalBiqSolarSolarMineralogy} it becomes clear that turbulent mixing has a significant effect on the mineralogy of dust in the disk midplane. For silicon (left panels in Fig. \ref{fig:32NonLocalBiqSolarSolarMineralogy}), the dominant carrier molecules are contained in magnesium-bearing minerals, such as forsterite (Mg$_2$SiO$_4$) and enstatite (MgSiO$_3$), while lizardite (Mg$_3$Si$_2$O$_5$(OH)$_4$) becomes the dominant silicate mineral at larger radial positions. In the local model, the transition from forsterite and enstatite towards lizardite occurs abruptly around $r=0.7$ AU, while the transition is very gradual in the nonlocal case. Here, more than 10 \% of the dust consists of lizardite at $r=0.4$ AU, while forsterite and enstatite also co-exist with lizardite exterior to $r=0.7$ AU. Lizardite overtakes the former two as the main silicate mineral between $r=0.42$ and $r=0.55$ AU. A similar behaviour occurs for potassium, calcium and aluminium-bearing condensates at lower molecular abundances. Individual species containing these elements usually comprise no more than a few mol \% of the solid phase. 

For the iron-bearing phases (right panels in Fig. \ref{fig:32NonLocalBiqSolarSolarMineralogy}), elemental iron constitutes the main reservoir for iron in the inner regions, with a gradual transition towards iron sulfide and abrupt change to fayalite (Fe$_2$SiO$_4$) at $r=0.3$ AU and another transition of fayalite into magnetite (Fe$_3$O$_4$) at $r=0.5$ AU as the main iron-bearing species. Again, when considering nonlocal disk processing, transitions in condensate compositions for iron-bearing phases are gradual instead of abrupt. The transition from metallic iron to iron sulfide as the main iron-bearing species occurs at $r=0.3$ AU in the local case. This shifts outward in the nonlocal case towards $r=0.4$ AU. Striking in the nonlocal scenario is the generally lower percentage of fayalite and magnetite at any given radius. Both minerals are encountered at lower abundances and spread over a larger range of radial positions in the nonlocal case.

For a nebular elemental composition akin to the abundances in HIP 43393's photosphere, which is more sulfur-rich than the Sun (Table \ref{tab:abundances}), Fig. \ref{fig:32NonLocalBiqSolarStar4Mineralogy} shows that the mineralogy of the condensates changes considerably. This is particularly true for the nature and abundance of iron-bearing species, where iron is almost exclusively incorporated in iron sulfide and pyrite (FeS$_2$), while fayalite and magnetite are not formed. In the local case, the transition from metallic iron into iron sulfide is more abrupt than for the Solar composition and is located at approximately $r\eqsim 0.13$ AU. In addition, a transition from iron sulfide to pyrite occurs at $r=0.6$ AU due to the higher availability of sulfur. For the silicates, the transition from olivine and pyroxene towards lizardite occurs at the same location as for the Solar composition, whereas the lack of fayalite results in more pyroxene compared to the local condensation sequence with Solar composition. 

In the nonlocal scenario, mineralogical transitions are again smoothed by turbulent diffusion, allowing for the coexistence of significant amounts of metallic iron, iron sulfide and pyrite at almost every radial distance interior to 1.5 AU. 

Despite the inclusion of several sulfates in GGCHEM, no appreciable amounts of these sulfates were found to form in any of the condensation sequences for both compositions considered in this work. Furthermore, additional tests revealed no compositional trends as a function of the size of aggregates in which monomers were embedded, suggesting that dust is collisionally well-mixed.

\subsection{Effects of devolatilization}
\label{sec:4.3.3}

\begin{figure}
    \centering
    \includegraphics[width=.48\textwidth]{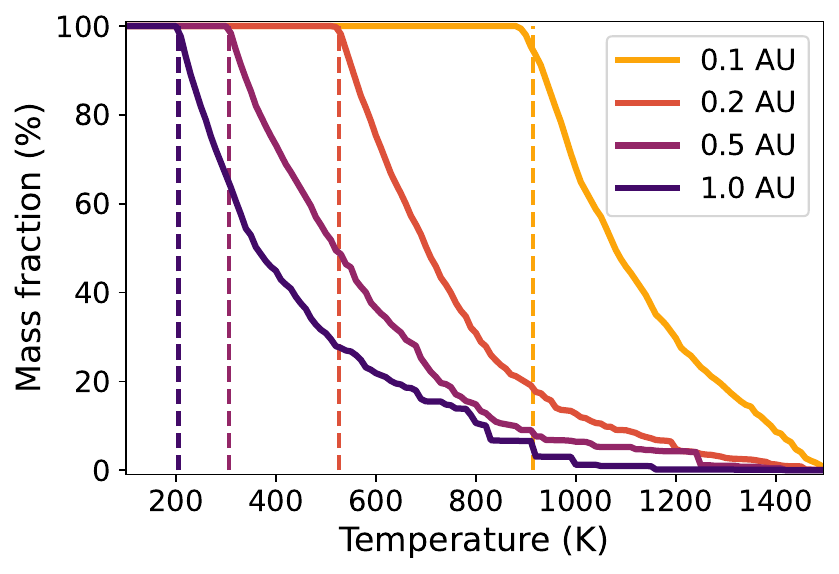}
    \caption{The fraction of the dust mass in the disk midplane at $r=0.1, 0.2, 0.5$ and $1$ AU that has been subjected to at least the given temperature due to dynamical transport towards warmer regions. The dashed lines indicate the dust temperature at the position of the local dust population.}
    \label{fig:33NonLocalBiqSolarMaxTemperatureFraction}
\end{figure}
\begin{figure*}
    \centering
    \includegraphics[width=\textwidth]{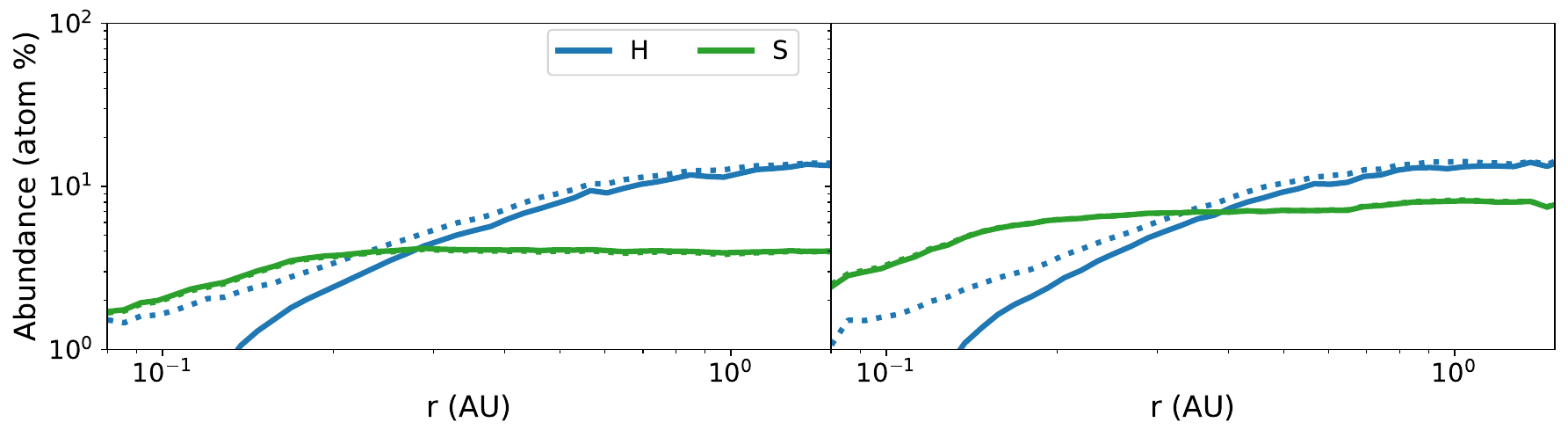}
    \caption{Radial elemental abundance of hydrogen and sulfur with (solid line) and without devolatilization (dotted line) for the model with Solar (left panel) and non-Solar composition (right panel).}
    \label{fig:33Devolatilization}
\end{figure*}

Nonlocal disk processes result in significant redistribution and mixing of material throughout the inner 1.5 AU of the disk midplane. This efficient radial and vertical mixing of monomers subjects monomers to a wide range of temperatures and pressures. This can result in dust monomers experiencing temperatures significantly above the temperatures at which they formed, which may result in chemical alteration (e.g. annealing or aqueous alteration) or even evaporation of minerals that constitute the refractory component of monomers. This may result in the loss of CHNOS from the planetesimal-forming dust. Fig. \ref{fig:33NonLocalBiqSolarMaxTemperatureFraction} depicts the maximum temperatures experienced by dust mass, probed by the local monomers at different radial positions. It becomes clear that extreme temperatures are experienced by a fraction of the local dust population at any radial position. The fraction of dust mass that is subjected to extreme temperatures decreases as a function of radial position. For example, temperatures above 1000 K have been experienced by only $\sim 1\%$ of the dust monomers visiting $r=1$ AU, whereas at $r=0.1$ AU, $\sim 65$ \% of the monomers have experienced temperatures above 1000 K.  

Many condensates in GGCHEM that are assigned to individual monomers are refractory. For example, silicates such as enstatite and forsterite and iron-rich condensates such as iron sulfide, magnetite or iron remain stable as a condensate at temperatures well beyond 1000 K. However, some condensates are not chemically inert when subjected to temperatures significantly higher than the temperatures at which the condensate originally formed under equilibrium conditions. Some of these condensates, such as hydrated silicates and sulfides can contain a significant fraction of the solid-phase CHNOS budget. We here aim to estimate an upper limit to the loss of CHNOS from the solid phase upon heating. Since no appreciable amounts of carbon- or nitrogen-bearing condensates have formed for both compositions (Fig. \ref{fig:32NonLocalBiqSolarSolarMineralogy} and Fig. \ref{fig:32NonLocalBiqSolarStar4Mineralogy}), we here focus on dehydration and desulfurization of minerals through thermal decomposition. 

The main mineral undergoing dehydration is lizardite, which starts undergoing thermal decomposition when subjected to temperatures above 773 K \citep[e.g.][]{akai_t-t-t_1992, nakamura_yamato_2006}. Although the thermal decomposition of lizardite is a multi-staged process as a function of temperature, we here consider the following reaction which comprises the full dehydration of lizardite \citep[]{akai_t-t-t_1992}{},
\begin{align}
\label{eq:dehydration}
    \Mg_3\Si_2\O_5(\O\H)_4\,\s\Rightarrow \Mg_2\Si\O_4\,\s + \Mg\Si\O_3\,\s + 2\H_2\O\,\g.
\end{align}
We assume reaction \ref{eq:dehydration} occurs proportionally to the fraction of monomers that have experienced a temperature above 773 K. This approach provides an upper limit on the dehydration of lizardite.

Under low-pressure conditions, it has been found that pyrite also remains stable up to 773 K, after which it decomposes into pyrrhotite (Fe$_{1-x}$S) and gaseous sulfur (S$_2$), with $x$ gradually decreasing towards zero for temperatures above $\sim1000$ K, where iron sulfide is formed \citep{xu_thermal_2019}. To provide an upper limit for desulfurization, we here assume that all pyrite decomposes into iron sulfide via
\begin{align}
\label{eq:desulfurization}
    2\,\Fe\S_2\,\s\Rightarrow2\,\Fe\S\,\s+\S_2\,\g
\end{align}
at a temperature of 773 K. The calculations shown in Fig. \ref{fig:33NonLocalBiqSolarMaxTemperatureFraction} can be performed at any grid cell in the background disk to calculate the fraction of the solid phase material in the disk midplane that has been exposed to temperatures higher than required to trigger the reactions shown in Eq. \ref{eq:dehydration} and Eq. \ref{eq:desulfurization}. We assume that the reactions occur instantaneously and no additional new condensates form except through the above reactions. This allows the effects of dehydration and desulfurization to be calculated via a combination of the mineral abundances from Fig. \ref{fig:32NonLocalBiqSolarSolarMineralogy} and Fig. \ref{fig:32NonLocalBiqSolarStar4Mineralogy} and temperature data calculated as in Fig. \ref{fig:33NonLocalBiqSolarMaxTemperatureFraction}. 

The effects of thermal decomposition on the radial elemental abundances of hydrogen and sulfur are shown in Fig. \ref{fig:33Devolatilization}. For both elemental compositions, the effects of dehydration on the elemental abundance of hydrogen are significant. Between $r=2$ and $r=0.15$ AU, hydrogen is gradually removed from the condensates as an increasing fraction of monomers have been exposed to temperatures higher than 773 K at smaller radial positions. For sulfur, the effects of desulfurization appear to be limited. This can mainly be attributed to the absence of pyrite in the disk regions where a significant fraction of monomers has been exposed to temperatures in excess of 773 K. For Solar elemental abundances, no pyrite was found to form as a condensate, while for the non-Solar composition, sulfur is primarily incorporated in iron sulfide in the warmer disk regions, with only small amounts of pyrite being present (Fig. \ref{fig:32NonLocalBiqSolarStar4Mineralogy}). 

\subsection{Planetesimal elemental abundances}
\label{sec:4.3.4}

\begin{figure*}
    \centering
    \includegraphics[width=\textwidth]{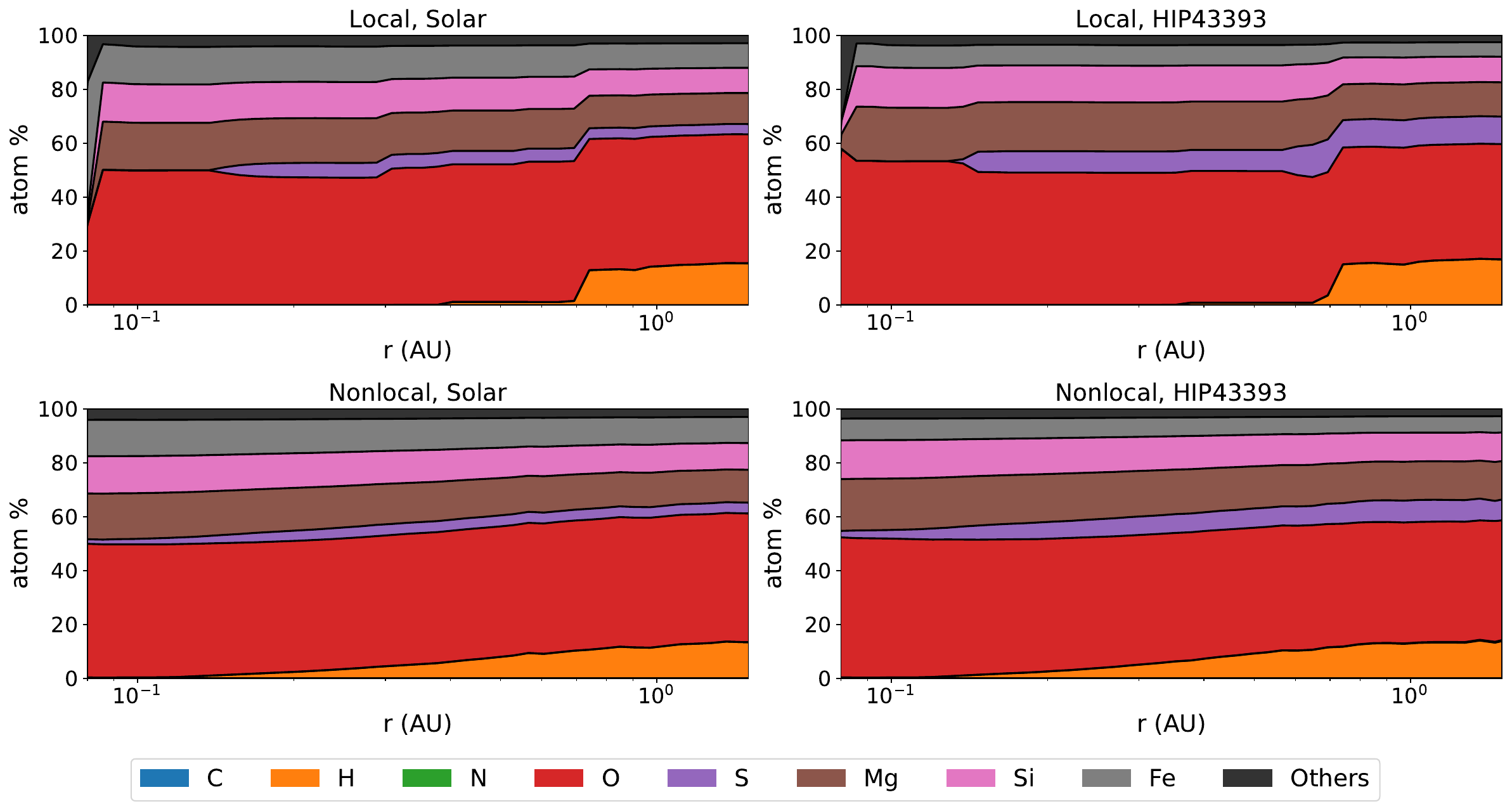}
    \caption{Radial behaviour of the elemental composition of dust for local equilibrium condensation (top row, inital conditions) and averaged over nonlocal dust monomers (bottom row, after 100 kyr), for condensation from a nebula with Solar and HIP 43393 composition (left and right column, respectively). Since we focus on the disk inside $\sim 1.5$~au, the minerals do not contain notable amounts of C and N.}
    \label{fig:34NonLocalBiqElementalAbundances}
\end{figure*}
\begin{figure*}
    \centering
    \includegraphics[width=\textwidth]{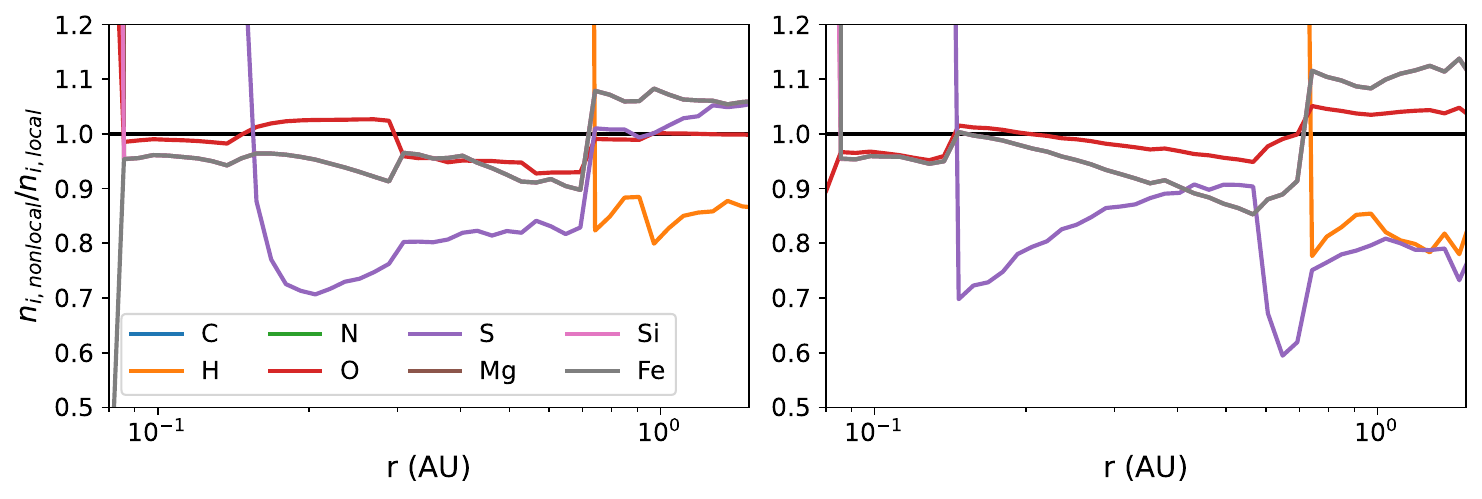}
    \caption{Comparison of the nonlocal and local radial behaviour of elemental atomic percentages $n_i$ (in at\%) for elemental species $i$. The left and right panels denote this comparison for the Solar and HIP 43393 composition, respectively. Since we focus on the disk inside $\sim 1.5$~au, the minerals do not contain notable amounts of C and N.}
    \label{fig:34ElementalDifferences}
\end{figure*}
Considering the mineralogies from Sect. \ref{sec:4.2.2}, accounting for the devolatilization of lizardite and desulfurization of pyrite gives rise to the elemental compositions shown in Fig. \ref{fig:34NonLocalBiqElementalAbundances}. It becomes clear that turbulent mixing efficiently transports solid-phase hydrogen and sulfur into disk regions previously depleted in these elements in the solid phase, enhancing the hydrogen and sulfur budget. Furthermore, we note that no carbon or nitrogen is present in the solid phase at any radial position for both compositions, suggesting that both elements are likely entirely incorporated in volatile molecules for the combinations of compositions, pressures and temperatures considered. Moreover, turbulent mixing results in the smoothing of the rather abrupt changes in elemental composition in the local models, which are typically associated with mineralogical transitions. An example is the formation of lizardite, resulting in a sudden increase of the amount of hydrogen at $r\eqsim 0.7$ AU. A similar behaviour can also be seen for sulfur for the transition of metallic iron into iron sulfide for both compositions and iron sulfide into pyrite for HIP 43393.

For both the Solar and HIP 43393 composition, the inward transport of hydrogen can be attributed to lizardite remaining stable to higher temperatures than the temperatures in the disk region where the lizardite forms from equilibrium condensation. Specifically, lizardite does not form under equilibrium condensation conditions interior to $r=0.7$ AU. However, the midplane gas and dust temperatures are approximately 250 K at this location, while lizardite remains stable to 773 K, a temperature that is only reached in the disk midplane inside $r=0.12$ AU. The gradual decrease in hydrogen percentage between $r=0.7$ and $r=0.12$ AU can be attributed to the gradually decreasing number of monomers containing lizardite towards smaller $r$ (see Fig. \ref{fig:32NonLocalBiqSolarSolarMineralogy} and Fig. \ref{fig:32NonLocalBiqSolarStar4Mineralogy}), and the increasing number of monomers having experienced higher temperatures due to excursions towards regions that are warmer than 773 K. The latter can be either due to radial excursions to smaller $r$ or vertical excursions towards the disk surface.  

Similar mechanisms apply to the inward transport of sulfur, where the inward transport of iron sulfide results in a significantly higher atomic percentage of sulfur in the innermost disk regions. For a Solar composition, the sulfur percentage at $r=0.1$ AU increased from 0.1 to 2.0 at\%, while for the composition of HIP 43393, the sulfur percentage was enriched from 0.1 to 3.1 at \%. The presence of sulfur at these small radial positions can be attributed to the presence of iron sulfide, while in the local models, almost no sulfur is present since iron forms as a metal at $r=0.1$ AU. For HIP 43394, pyrite also contributes to the inward transport of sulfur. However, pyrite decomposes into iron sulfide at $r=0.12$, exacerbating the gradual decrease in sulfur percentage towards smaller radial positions. 

In Fig. \ref{fig:34ElementalDifferences}, we explore the difference between the local and nonlocal compositions by radially calculating the difference of the elemental atomic percentage in both cases for the Solar composition and composition of HIP 43394 shown in Fig. \ref{fig:34NonLocalBiqElementalAbundances}. This comparison reveals that the transport of hydrogen and sulfur not only enriches the regions where less hydrogen- and sulfur-bearing minerals form in the local case but also depletes the disk regions where hydrogen- and sulfur-bearing minerals were readily forming locally. This originates from the fact that these regions act as source regions for hydrogen- and sulfur-rich material in the nonlocal case. The depletion is therefore the largest at locations where large gradients in elemental composition are present. For the Solar composition, this results in a sulfur depletion up to $29\%$ at $r=0.2$ AU for the local composition, and up to $41\%$ at $r=0.65$ AU for the HIP 43393 composition. This is expected since these positions correspond to the location of the transition from metallic iron to iron sulfide in the Solar mineralogy, and the transition from iron sulfide to pyrite in the HIP 43393 composition (Fig. \ref{fig:32NonLocalBiqSolarSolarMineralogy} and Fig. \ref{fig:32NonLocalBiqSolarStar4Mineralogy}). A similar argument applies to hydrogen and the transition from enstatite and forsterite into lizardite, with a hydrogen depletion of $\sim 20\%$ around $r=0.7$ AU being reached for both the Solar and HIP 43393 composition. 

Surprisingly, the effects of nonlocal processing on the amounts of magnesium, silicon and iron fully coincide in Fig. \ref{fig:34ElementalDifferences}. This can be attributed to the fact that the effects of devolatilization processes on these elements are small compared to hydrogen and sulfur, which means that the changes in the atomic percentage of these species are driven by the influx of hydrogen and sulfur. Due to the higher percentage of hydrogen and sulfur at smaller radial positions, the percentages of the solid phase components containing magnesium, silicon and iron become smaller, resulting in up to 10 at\% less of these species at $r\lesssim0.7$ AU. At $r\gtrsim0.7$ AU, these elements tend to become enriched due to the depletion of hydrogen and sulfur in these regions. Therefore, the enrichment of magnesium, silicon and iron in this region is also larger for HIP 43393 due to the depletion of pyrite in addition to lizardite. For oxygen, the most abundant element for both compositions, this effect also applies due to its presence in minerals. However, the transport of lizardite can offset this loss since hydrogen is incorporated as hydroxyl groups, such as between $r=0.15$ AU and $r=0.3$ AU for the Solar composition. Due to the outflux of lizardite from the region exterior to $r=0.7$ AU, oxygen does not follow the enrichment in magnesium, silicon and iron in the nonlocal scenarios, but remains rather close to the oxygen atomic percentage of the local scenarios for both stellar compositions. This combined effect results in oxygen being the element least affected by nonlocal processes.
    
    \section{Discussion}
    \label{sec:3.4}

\subsection{Solar System context}
\label{sec:4.4.1}

Overall, efficient nonlocal disk processing dominated by turbulent diffusion results in the homogenization of planetesimal-forming dust in the disk midplane. In the Solar System context, planetesimals in the inner Solar System are thought to have formed in the first $1\e{5}$ yr to $3\e{5}$ yr after the formation of calcium-aluminium inclusions \citep[CAI, e.g.][]{kruijer_protracted_2014, lichtenberg_bifurcation_2021}. This timescale is comparable to the timescale considered in the simulations in this work. Thus nonlocal disk processing can result in considerable compositional homogenization of refractory material in the inner Solar System, also with the inclusion of collisional processing. Radial mixing of dust in the protosolar nebula is thought to be necessary to explain the presence of high-temperature products such as crystalline silicates and calcium-aluminium inclusions in asteroids and comets \citep[e.g.][]{ciesla_residence_2011, aguichine_rocklines_2020, jang_spatial_2024, woitke_cai_2024} and the inheritance of interstellar ices by asteroids and comets \citep[e.g.][]{bockelee-morvan_turbulent_2002, bergner_ice_2021}. This homogenization is also consistent with, the observed variations in the water content of chondrites, which exhibit a gradual increase as a function of heliocentric distance from water-poor ordinary chondrites ($\lesssim 1$ wt\% H$_2$O) to water-rich CI chondrites ($\sim 14$ wt\% H$_2$O) \citep[][]{piani_origin_2021, jones_meteorites_2024}. Carbonaceous chondrites typically also contain significant amounts of carbon (up to $\sim 4$ wt \%), whereas our model depicts no solid-phase carbon. However, carbon found in primitive chondrites suggests that carbon may be inherited from presolar grains \citep[e.g.][]{huss_presolar_1995, huss_presolar_2003, christ_open_2024}. In addition, the parent bodies of carbonaceous chondrites are thought to have accreted much further out than the region studied here \citep[beyond Jupiter, e.g.][]{kruijer_age_2017}. Altogether these lines of Solar System evidence support large-scale non-local disk processing before the formation of planetesimals. It is also clear, that the equilibrium condensation assumed in this study neglects the inheritance of presolar material, which can have significant effects on the estimates obtained for specific element abundances, such as carbon. 
We also note that for the silicates, our results are in line with the abundances of fayalite in olivine and the olivine to pyroxene ratio derived for near-infrared spectra of near-Earth ordinary chondrite asteroids \citep[][]{sanchez_population_2024}. 

\subsection{Implications for planetesimal composition}
\label{sec:4.4.2}

Assuming planetesimal formation through a non-violent mechanism, such as gravitational collapse over a few orbital timescales triggered by e.g. the streaming instability \citep[][]{youdin_streaming_2005, blum_evidence_2017, nesvorny_trans-neptunian_2019, visser_radial_2021}, the resulting initial planetesimal mineralogy could be fully inherited from the mineralogy of the constituent dust. This means that the predicted dust mineralogy in this work can be projected directly onto the initial compositions of planetesimals that would form from this dust. Based on this notion, nonlocal disk processing tends to homogenize refractory dust composition in the inner disk. This implies that the initial compositions and elemental budgets of planetesimals forming at different radial positions interior to the water ice line become more homogenized as time progresses. 

We also note that the extent of this effect depends on the turbulence strength, which was assumed to be characterized by a single value $\alpha=10^{-3}$. This value is consistent with values required to explain turbulence-driven accretion \citep[e.g.][]{delage_steady-state_2022, trapman_observed_2020, rosotti_empirical_2023}. Turbulence originating from the magneto-rotational instability (MRI) \citep[][]{balbus_powerful_1991} has been widely studied as the driver of accretion, although large regions of the disk may be dynamically "dead-zones" where this mechanism is ineffective \citep[e.g.][]{trapman_observed_2020, lesur_hydro-_2023}. These values of $\alpha$ are also high when compared with values derived from millimetre continuum observations of dust settling, which may range from $\alpha \sim 10^{-3}$ down to values below $\alpha\sim10^{-5}$ \citep[e.g.][]{pinte_dust_2016, ueda_impact_2021, doi_estimate_2021}. However, we also note that these studies primarily focus on the outer regions of Class II disks, while settling in the inner regions of Class I disks is less well studied \citep[e.g.][]{rosotti_empirical_2023}. Thus, various lines of evidence suggest that the inner protoplanetary disks may in reality be less turbulent than assumed in this work, which would limit the transport of dust via turbulent diffusion. Since the Stokes numbers of dust grains are well below St$\eqsim 10^{-3}$ in the inner disk of our disk model \citep[][]{oosterloo_effect_2024}, aerodynamic processes are likely even less efficient in affecting dust grain composition through the transport of dust on the planetesimal formation timescale. Altogether we would expect compositional gradients in the dust and planetesimal composition of a less turbulent disk to be steeper.

The subsequent evolution of planetesimals involves heating of the planetesimal interior by the decay of the radioactive isotopes $^{26}$Al and $^{56}$Fe which can result in the devolatilization, melting and differentiation of planetesimals \citep[e.g.][]{macpherson_distribution_1995, hevey_model_2006}. These processes can result in the loss of the volatile material, such as water, from the planetesimal, depending on planetesimal mass and abundance of radioactive isotopes \citep[e.g.][]{lichtenberg_water_2019}. This suggests that part of the hydrogen transported towards smaller radial positions can still be lost from the planetesimal interior during thermal metamorphism. However, considering that most of the additional hydrogen is present as water, any water molecules liberated from the hydrated minerals could oxidize surrounding material while escaping \citep{mcsween_oxidation_1992, lewis_phosphate_2016}. From the results from Fig.~\ref{fig:34ElementalDifferences} we expect that the effects on overall redox conditions in the inner disk solids remain limited since the gain in solid-phase oxygen due to the influx of hydrated minerals comes at the loss of other oxygen-bearing minerals.

In none of the 32000 condensation sequences run in this work (16000 monomers of each composition) sulfates have formed, despite their implementation in the GGCHEM code. Evidently, sulfates do not form under equilibrium conditions at $r<1.5$ AU for a Solar composition at these pressures and are also not expected to have formed in this region around HIP 43393. This is to some extent expected since sulfate minerals on Earth generally form under redox conditions that are considerably more oxidizing \citep[e.g.][]{jugo_experimental_2004} than those thought to have prevailed during planet formation in the inner Solar System \citep[e.g.][]{rubie_heterogeneous_2011, steenstra_constraints_2016, doyle_oxygen_2019}{}{}. Nevertheless, sulfate minerals (predominantly gypsum) have been found in CC and K3 chondrite and enstatite achondrite meteorites, and are commonly thought to be associated with either terrestrial or localized plantesimal aqueous alteration \citep{zolensky_aqueous_1988, brearley_CI_1992, buseck_matrices_1993, brearley_nature_1989, izawa_que_2011, airieau_planetesimal_2005, suttle_aqueous_2021}. Isotopic signatures in CC bodies suggest that they could have formed from a separate dust reservoir in the outer Solar System, where water ice is more prevalent \citep[e.g.][]{kruijer_great_2020}{}{}. We do note that the GGCHEM code currently does not take into account the formation of gypsum (CaSO$_4$) as a condensate. However, Table \ref{tab:abundances} shows that calcium is considerably less abundant than iron and magnesium, elements that have been included as sulfates. Also, the abundance of calcium is approximately 6 to 21 times lower than the abundance of sulfur for the Solar and HIP 43393 compositions respectively. Therefore, even if efficient formation of gypsum would consume the entire calcium elemental budget, sulfates would only comprise a limited fraction of the sulfur elemental budget. Altogether our results do not support the formation of sulfate minerals before planetesimal formation.

\subsection{Model limitations}
\label{sec:4.4.3}

Although our model provides a fully coupled description of the effects of dynamical and collisional processing on individual units of dust mass, the composition of dust is derived from equilibrium condensation conditions, based on the minimization of the total Gibbs free energy. This means that all chemical processes that give rise to the thermodynamically most favourable solid phase state are assumed to happen on timescales much shorter than the simulation timescale of 100 kyr. This is likely not the case towards lower temperatures and would require the incorporation of non-equilibrium chemistry. Although adsorption and condensation timescales become shorter towards lower temperature, this is not true for the chemisorption and annealing processes that give rise to the formation of e.g. phyllosilicates or crystalline silicates \citep[][]{dangelo_water_2019, herbort_atmospheres_2020, oosterloo_shampoo_2023, jang_spatial_2024, woitke_cai_2024}. Although the rearrangement of minerals in the solid phase does not directly affect their CHNOS budgets, newly formed species could incorporate certain elements in mineral phases that are more or less refractory, which could affect the loss of these elements through decomposition and evaporation processes. We expect the elemental budgets of hydrogen and oxygen to be most affected by these additional chemical processes due to the incorporation of water in hydrated minerals. For example, these effects could result in lower amounts of hydrogen and oxygen in regions where our current model labels monomers with compositions where silicates exist in the form of lizardite (mostly outside $r=0.7$ AU as this lizardite likely results from the gradual hydration of enstatite and forsterite). However, more hydrogen and oxygen could be present in the disk regions where enstatite and forsterite are present as minerals, where gas-phase water could result in the gradual hydration of these minerals within the 100 kyr timescale considered in this work \citep[e.g.][]{dangelo_water_2019}. 

A similar argument on chemical equilibrium timescales applies to our approach to devolatilization. We account for the effects of the thermal decomposition of lizardite and pyrite by assuming instantaneous, complete decomposition above a certain critical temperature. This approach leads to a lower limit on the amounts of sulfur and hydrogen retained in these mineral phases. However, due to the sensitivity of chemical timescales on temperature and irradiation and thus location in the disk \citep[e.g.][]{woitke_radiation_2009, cuppen_grain_2017}, to improve estimates for the behaviour of devolatilization it is desirable to include an adsorption/desorption model for the net gain and loss of molecules akin to the framework for volatile molecules included in \cite[][]{oosterloo_shampoo_2023}, coupled to an internal chemical evolution model for monomer compositions. 

If the disk undergoes continuous accretion in the class\,I phase ($< 0.2$~Myr), fresh material could be delivered interior to the centrifugal radius and participate in the collisional and dynamical mixing of dust inside $\sim 5$~au \citep{hueso_evolution_2005}. Such infall --- along with outward transport --- could also affect the calculation of our initial equilibrium composition. \citet{Min2011} and \citet{woitke_cai_2024} show that the inner disk thermostat mechanism (dust stability) is by itself capable of altering and equilibrating the mineralogy of the accreted material and we may not even require outbursts to sublimate the solids in the inner disk. The efficient collisional recycling of monomers inside a few au \citep[e.g.][]{oosterloo_shampoo_2023} can ensure that the annealing process occurs on timescales much shorter than 100~kyr. So, while we neglect the infall of fresh material, the effect on our initial conditions may be limited.

    \section{Conclusions and outlook}
    \label{sec:3.5}

In this work, we explored the effects of nonlocal disk processing, coupled with collisional processing and equilibrium condensation in a young, massive, static class I disk on the refractory CHNOS budgets of dust inside $r=1.5$ AU. We considered elemental abundances consistent with the photospheric abundances of the Sun and HIP 43393, a star with a super-solar S/Fe ratio. Moreover, we considered a simple devolatilization scheme for hydrated minerals and pyrite. 

Efficient turbulence-driven diffusion was found to result in the entire inner disk becoming well-mixed within 10$^5$ yr (Fig. \ref{fig:31NonLocalBiqSolarOriginDiagram} and Fig. \ref{fig:31NonLocalBiqSolarOriginDiagramSnapshots}). This gives rise to mineralogies of midplane dust that are radially considerably more mixed than predicted from local condensation under equilibrium conditions. For both compositions studied, mineralogical transitions are considerably smeared out with respect to fully local dust that formed under equilibrium condensation conditions. The elements whose midplane abundances are affected the most are hydrogen and sulfur. For hydrogen, this is a consequence of the inward transport of hydrated minerals such as lizardite, while for sulfur, this originates from the inward diffusion of iron sulfide and pyrite. 
 
Due to the efficient radial and vertical turbulent mixing of dust monomers, reflecting individual units of dust mass, dust in the disk midplane at a given radial position constitutes material that has been subjected to a wide range of maximum temperatures (Fig. \ref{fig:33NonLocalBiqSolarMaxTemperatureFraction}). For example, at $r=1$ AU, $\sim 1\%$ of the dust mass has experienced a temperature above 1000 K, increasing to $\sim 65\%$ of the dust mass at $r=0.1$ AU. This thermal processing can result in significant loss of elemental hydrogen inside $r\lesssim 0.3$ AU (Fig. \ref{fig:33Devolatilization}), transported in the form of hydrated minerals from larger radial distances. Radially transported sulfur appears to be more resistant to higher temperatures, with no significant sulfur loss resulting from the thermal decomposition of pyrite for the composition of HIP 43393. 

Altogether nonlocal disk processing could lead to significant compositional homogenization in the dust situated in the disk midplane interior to $r=1.5$ AU. This mineralogy is more enriched in hydrogen at smaller radial positions due to the transport of hydrated minerals (Fig. \ref{fig:34ElementalDifferences}), which may result in more aqueous alteration during planetesimal evolution. Future studies are recommended to implement a more comprehensive model for the time-dependent chemical evolution of the mineralogy of dust prior to planetesimal formation.   
    
    \section*{Acknowledgements}
    This work is part of the second round of the Planetary and Exoplanetary Science Network (PEPSci-2), funded by the Netherlands Organization for Scientific Research (NWO). We also thank Peter Woitke for implementing the thermodynamic data of sulfates in the GGCHEM code and Rob Spaargaren and Tim Lichtenberg for insightful discussions.


\bibliographystyle{aa.bst}
\bibliography{references_additions.bib}

\begin{appendix}

\section{Interpreting individual monomer trajectories}
\label{sec:4.AA}

\begin{figure*}[h!tbp]
    \centering
    \includegraphics[width=\textwidth]{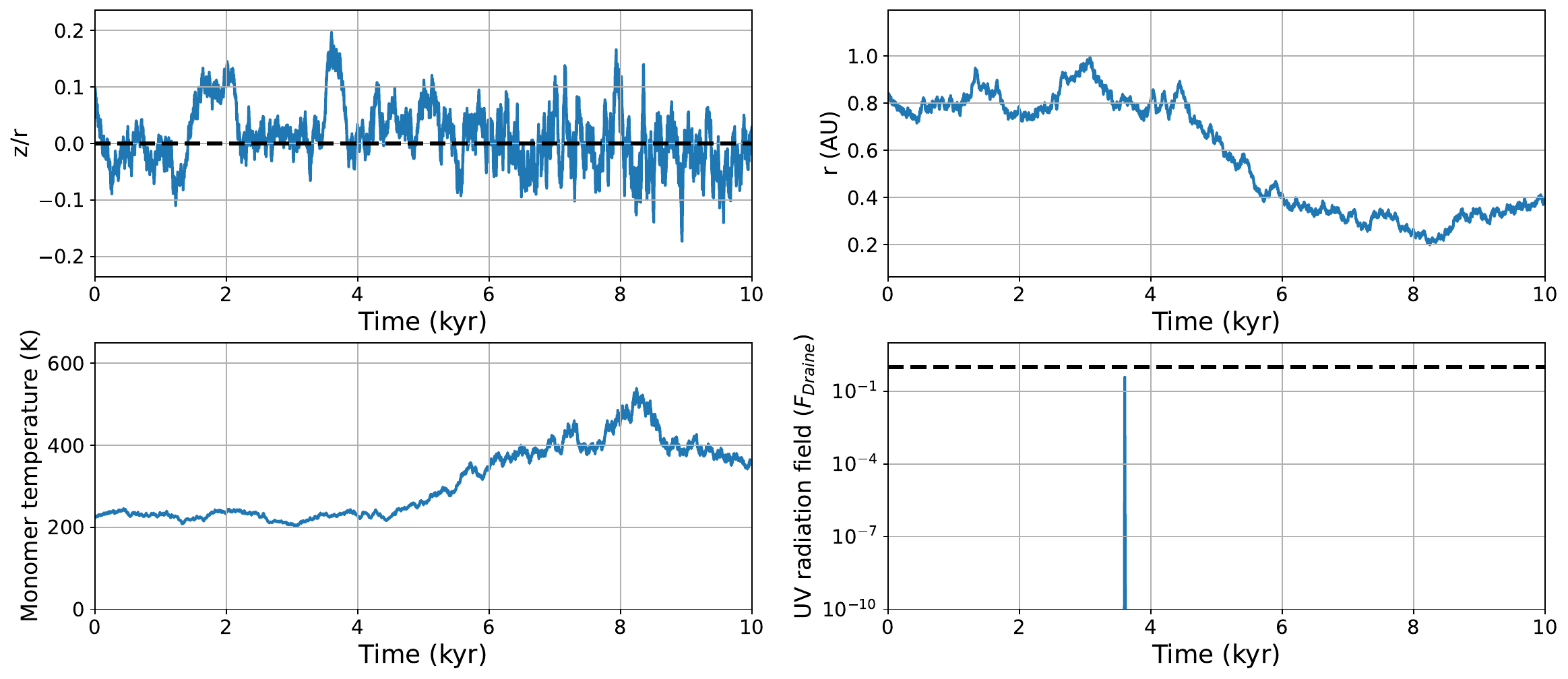}
\end{figure*}
\begin{figure*}[h!tbp]
    \centering
    \includegraphics[width=\textwidth]{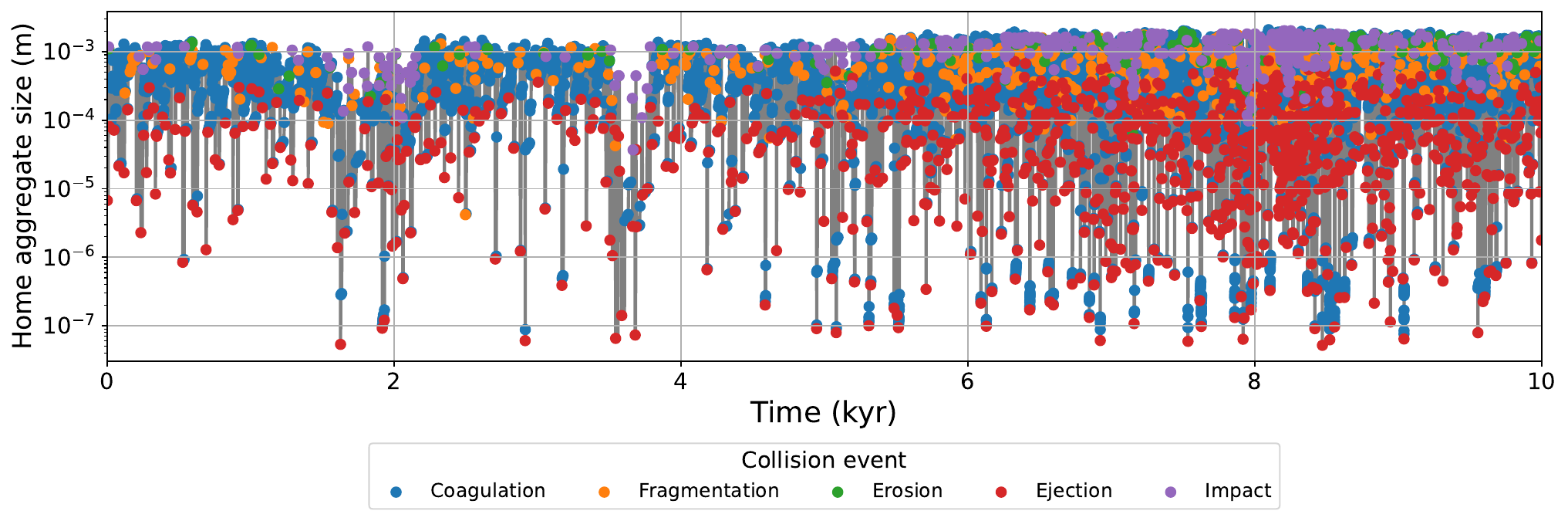}
    \caption{Example of the first $10^4$ years of a monomer trajectory calculated with SHAMPOO in this work. This monomer originates from $r_0\approx 0.8$ AU and $z_0/r_0\approx 0.1$ and is undergoing dynamical and collisional processing. All quantities are shown as a function of time. The upper left and right panels show the vertical and radial monomer positions and the centre left and right the resulting monomer temperature and UV radiation field in the local environment. The bottom panel displays the size of the aggregate in which the monomer is embedded, with each dot indicating an individual collision event. The black dashed line in the upper left panel indicates the disk midplane, while the black dashed line in the centre right indicates the radiation field strength of $1F_\text{Draine}$.}
    \label{fig:AAExampleMonomer}
\end{figure*}

In this work, we utilize the average behaviour of 16 000 individual dust monomers to make inferences about the average dust refractory composition at various locations in the disk. In this appendix, we provide a short overview of the dynamical and collisional evolution of an individual dust monomer in SHAMPOO. We do emphasize that the stochastic nature of individual monomer trajectories means that to study the implications of the behaviour of monomers for local dust populations, averaging over the behaviour of many monomers is necessary. In this work, we focussed on the dynamical behaviour of monomers interior to $r=1.5$ AU. The average amount of ice on dust monomers was found to become negligible compared to the total dust monomer mass between $r=1.5$ to $r=2.0$ AU, which is why ice processing was ignored in this study. For evolutionary trajectories of individual monomers at larger radial distances including ice processing, we refer the interested reader to \cite{oosterloo_shampoo_2023}, while the implications of nonlocal behaviour on ice processing have been studied in \cite{oosterloo_effect_2024}. 

Fig. \ref{fig:AAExampleMonomer} shows the trajectory of one of the monomers utilized in this work. For clarity, only the first 10 kyr of the full 100 kyr trajectory are shown. In Sect. \ref{sec:4.2.1} we noted that turbulence-driven diffusion is the dominant transport process, which makes the spatial trajectories of individual monomers highly stochastic. Vertically, this behaviour allows the monomer to make a single excursion to $|z/r|\eqsim0.2$, and many excursions to $|z/r|\eqsim 0.1$. Radially, the monomer initially remains between $r=0.8$ and $r=1.0$ AU, until it undergoes an episode of inward diffusion between 4 kyr and 6 kyr, where it ends up at radial positions between $r=0.4$ and $r=0.6$ AU from 6 kyr to 10 kyr. This vertical and radial transport has significant effects on the monomer temperature, the local UV radiation field and the aggregate size in which the monomer is located. The effects of vertical excursions on the monomer temperature appear to be limited, while the vertical excursion between 3 and 4 kyr appears to result in the only occasion where the monomer is exposed to notable amounts of UV radiation. The monomer temperature appears to follow changes in the radial distance, with the monomer temperature steadily increasing from 240 K at 4.0 kyr to 540 K around 8.2 kyr as a consequence of the gradual radial diffusion from $r\eqsim0.8$ AU to $r\eqsim0.2$ AU. 

Although the effects of the monomer aggregate size on transport are negligible interior to $r=1.5$ AU, the reverse is far from true. Fig. \ref{fig:AAExampleMonomer} shows that within the first 10 kyr of its evolution, the monomer has undergone many (more than 15 000) collisions, with the aggregate in which the monomer is embedded (the home aggregate) changing size during each collision. Here, we \sout{here} distinguish between coagulation and impact, which are the collision outcomes that result in a net growth of the home aggregate, and fragmentation, erosion and ejection, which result in a decrease in home aggregate size. The vertical excursions around 2 kyr and between 3 and 4 kyr are also visible in the collisional history, with the monomer becoming embedded in smaller aggregates on average and less coagulation taking place during these excursions, which can be explained by the lower number of available collision partners at higher $|z/r|$. The radial diffusion inwards after 4 kyr also has significant effects on the collisional history. Due to the higher gas and dust density at smaller $r$, collisions occur more frequently, with the monomer becoming mixed more frequently between small and large aggregates. Furthermore, aggregates appear to be able to approximately grow a factor 2 larger before a destructive collision outcome (fragmentation, erosion or ejection) results in a decrease in home aggregate size.

\end{appendix}

\end{document}